\documentclass[preprint,12pt]{elsarticle}

\usepackage{amssymb}
\usepackage{amsmath}

\journal{Earth and Planetary Science Letters}

\begin{document}

\begin{frontmatter}

\title{Incompatibility of iron in post-perovskite and the stability of basal magma oceans in super-Earths}

\author{Francis Dragulet\corref{cor1}}
\ead{francisdragulet@g.ucla.edu}
\author{Lars Stixrude}
\cortext[cor1]{Corresponding author}

\affiliation{organization={Department of Earth, Planetary, and Space Sciences, University of California, Los Angeles},
            addressline={595 Charles E. Young Drive East},
            city={Los Angeles},
            postcode={90095},
            state={California},
            country={USA}}

\begin{abstract}
Post-perovskite is expected to dominate much of the solid mantles of rocky exoplanets, yet iron partitioning between post-perovskite and silicate melt, which controls the compositional evolution and buoyancy of crystallizing magma oceans, is unconstrained at these pressures. We use first-principles molecular dynamics and thermodynamic integration to compute the Fe--Mg distribution coefficient $K_D$ between post-perovskite and (Mg,Fe)SiO$_3$ liquid at 150--600~GPa and 6000--10000~K. Iron is strongly incompatible in post-perovskite and becomes increasingly so with pressure. Combining $K_D$ with equations of state, we find that iron enrichment of residual liquid reverses the solid--liquid density contrast, causing post-perovskite to become buoyant. Basal magma oceans are therefore gravitationally stable in super-Earth exoplanets up to 4 M$_{\oplus}$.%
\end{abstract}


\end{frontmatter}

\section{Introduction}
\label{sec:intro}

Mass--radius relations of discovered exoplanets reveal a large population consistent with predominantly rocky compositions: likely a metallic iron-rich core surrounded by a silicate mantle \cite{dorn2015can,valencia2006internal}. Many of these planets are more massive than Earth and thus reach interior pressures far beyond those within Earth. Throughout most of Earth's lower mantle (25-136 GPa), MgSiO$_3$ adopts the bridgmanite structure. However, above $\sim$120$-$160~GPa, depending on temperature, bridgmanite transforms to post-perovskite \cite{murakami2004post,tateno2009determination}, with a density increase of $\sim$1--1.5\% across the transition \cite{komabayashi2008simultaneous}. Post-perovskite is expected to remain stable up to $\sim$500$-$700~GPa \cite{umemoto2017phase}, roughly corresponding to core-mantle boundary pressures of 4-5 M$_{\oplus}$ super-Earths. Post-perovskite may therefore be the dominant silicate phase through most of the mantles of super-Earths.

The early mantles of these rocky exoplanets likely experienced extensive melting. Accretionary impacts melt an increasing fraction of a planet as its mass increases, while gravitational energy released during core formation may itself be sufficient to cause global melting \cite{stixrude2014melting}. Intense stellar irradiation of close-in rocky planets and insulation by hydrogen-rich envelopes in the early stages of evolution can further promote mantle melting \cite{tang2025reassessing}. These effects suggest that deep or global magma oceans, and their subsequent crystallization, constitute an important stage in the evolution of these planets.

On Earth, crystallization of a molten mantle may have produced a basal magma ocean (BMO) that persisted for billions of years \cite{labrosse2007crystallizing,boukare2025solidification}. A central mechanism favoring such a reservoir is the incompatibility of iron in lower-mantle minerals \cite{nomura2011spin,braithwaite2022partitioning,dragulet2024partitioning,caracas2019melt}. As crystallization proceeds, iron partitions into the residual liquid, producing relatively buoyant crystals and a dense, iron-enriched melt that can remain gravitationally trapped at the base of the mantle. The presence and persistence of this silicate melt layer can substantially influence planetary structure and long-term evolution. For example, a BMO can modify the cooling history of the core \cite{auerbach2025thermal, lherm2024thermal}, concentrate incompatible and radiogenic elements into a long-lived deep reservoir \cite{labrosse2007crystallizing,deng2021deep}, and generate a magnetic field if it is sufficiently electrically conductive \cite{ziegler2013implications,soubiran2018electrical,stixrude2020silicate, dragulet2025electrical, nakajima2026electrical}. The formation of analogous reservoirs in super-Earths is therefore important for the subsequent thermal, chemical, and magnetic evolution of these planets.

Whether basal magma oceans can form in super-Earths is unknown because their magma oceans crystallize over a pressure range in which post-perovskite, rather than bridgmanite, is the relevant solid silicate phase. The key quantity is the Fe--Mg distribution coefficient between post-perovskite and (Mg,Fe)SiO$_3$ liquid, $K_D$, which controls how efficiently crystallization enriches the residual liquid in iron. Diamond-anvil cell experiments at 140--180~GPa find $K_D \approx 0.04$--$0.10$ \cite{nomura2011spin,tateno2014melting}, demonstrating strong iron incompatibility at the low-pressure end of the post-perovskite stability field. How this partitioning evolves across the large pressure range of super-Earth mantles, and how it affects crystal buoyancy, remain unconstrained.

Here we use first-principles molecular dynamics simulations and thermodynamic integration to compute $K_{D}$ between post-perovskite and (Mg,Fe)SiO$_{3}$ liquid at 150--600~GPa and 6000--10000~K, conditions spanning the mantles of super-Earths up to $\sim$4~$M_{\oplus}$.  We combine our computed distribution coefficient with equations of state of both phases to evaluate crystal buoyancy during magma ocean crystallization. We also use $K_{D}$ to estimate the depression of the liquidus of the iron-enriched residual liquid. We show that both effects favor the formation and longevity of basal magma oceans in super-Earths.

\section{Methods}
\label{sec:methods}

\subsection{Distribution coefficient from thermodynamic integration}
\label{sec:methods-kd}

The calculation of the distribution coefficient follows our recent study of iron partitioning in bridgmanite \cite{dragulet2024partitioning}. Briefly, the Fe--Mg distribution coefficient between post-perovskite (ppv) and (Mg,Fe)SiO$_{3}$ liquid (liq) is
\begin{equation}
K_{D} = \frac{X_{\mathrm{Fe}}^{\mathrm{ppv}}\,X_{\mathrm{Mg}}^{\mathrm{liq}}}
{X_{\mathrm{Fe}}^{\mathrm{liq}}\,X_{\mathrm{Mg}}^{\mathrm{ppv}}}
= \exp\!\left(-\frac{\Delta\tilde{\mu}_{R}}{k_{B}T}\right),
\label{eq:KD}
\end{equation}
where $X_{\mathrm{Fe}} = \mathrm{Fe/(Fe+Mg)}$ and $\Delta\tilde{\mu}_{R}$ is the chemical-potential change of the cation-exchange reaction
\begin{equation}
\mathrm{MgSiO_{3}\,(ppv)} + \mathrm{FeSiO_{3}\,(liq)}
=
\mathrm{MgSiO_{3}\,(liq)} + \mathrm{FeSiO_{3}\,(ppv)}.
\label{eq:rxn}
\end{equation}

The chemical potential of component $i$ is related to the Helmholtz free energy by \(\mu_i = \left(\partial F/\partial N_i\right)_{V,T,N_{j\neq i}} \), so the exchange chemical potential can be obtained from the free energy change of substituting Fe for one Mg atom  in each phase, \(\Delta\tilde{\mu}_{R} = \Delta F_{\mathrm{ppv}}-\Delta F_{\mathrm{liq}}. \) We compute each substitution free energy by thermodynamic integration along a coupling parameter $\lambda$ that interpolates linearly between the iron-free ($\lambda=0$) and iron-bearing ($\lambda=1$) Hamiltonians:
\begin{equation}
\Delta F =
\int_{0}^{1}
\left\langle \frac{\partial U}{\partial \lambda} \right\rangle_{\lambda} d\lambda
\approx
\frac{1}{2}
\left[
\left\langle \Delta U \right\rangle_{0}
+
\left\langle \Delta U \right\rangle_{1}
\right]
\label{eq:TI}
\end{equation}
where $\Delta U = U_{\mathrm{Fe}}-U_{\mathrm{Mg}}$ is the difference in total energy between the iron-bearing and iron-free systems evaluated on the same configuration, and angle brackets denote time averages over the corresponding molecular dynamics trajectory. The total energy $U = E - TS_{\mathrm{el}} - TS_{\mathrm{mag}}$ includes the electronic entropy $S_{\mathrm{el}}$ and the magnetic entropy $S_{\mathrm{mag}} = k_{B}\ln(m + 1)$, with $m$ the magnetic moment of the iron atom in Bohr magnetons. Because a single cation is exchanged, our calculations here yield the distribution coefficient in the dilute limit. However, we apply this dilute-limit coefficient at finite iron content because our previous calculations found that the iron substitution free energy is independent of $X_{\mathrm{Fe}}$ across the entire (Mg,Fe)SiO$_3$ join, for both bridgmanite and the same silicate liquid \cite{dragulet2024partitioning}.

\subsection{Simulation details}
\label{sec:methods-fpmd}

Our simulations are based on density functional theory, using the PBEsol exchange-correlation functional \cite{perdew2008restoring}, and projector augmented-wave method \cite{blochl1994projector,kresse1999ultrasoft}, as implemented in VASP \cite{kresse1996efficiency}. We perform Born--Oppenheimer molecular dynamics in the $NVT$ ensemble with periodic boundary conditions, a Nos\'e--Hoover thermostat, and a 1~fs timestep, with each trajectory run for 8--12~ps after equilibration. We assume thermal equilibrium between ions and electrons via the Mermin functional \cite{mermin1965thermal}, with the electronic temperature set equal to the thermostat temperature. Sampling the Brillouin zone at the $\Gamma$ point with a plane-wave basis cutoff of 600~eV converges the energy and pressure to within 4~meV/atom and 0.2~GPa, respectively. Prior to running simulations of post-perovskite, we adjust the lattice parameters of the orthorhombic cell at constant volume to ensure that the stress is hydrostatic. Liquid simulations are initialized by melting the crystalline structure at high temperature and equilibrating at the temperature of interest. Calculations on iron-bearing systems are spin-polarized and use the rotationally invariant DFT$+U$ scheme of Dudarev \cite{dudarev1998electron} with $U - J = 2.5$~eV, the value adopted in our previous work \cite{holmstrom2015spin,holmstrom2016spin,dragulet2024partitioning}. The simulation cell contains 24 MgSiO$_{3}$ formula units (120 atoms), with Fe substituted for one Mg atom in the iron-bearing runs.

\subsection{Spin state of iron}
\label{sec:methods-spin}

Over the large pressure range considered here, the local magnetic moment of iron is expected to change continuously with compression rather than remaining near a single high- or low-spin value. We therefore allow the magnetic moment to evolve freely during the molecular dynamics trajectories, rather than imposing moments of fixed magnitude. The time-averaged moment from our free-spin molecular dynamics simulations decreases from $\sim2.5~\mu_B$ near 150~GPa to nearly zero at the highest pressures explored (Fig.~\ref{figS1}a), with the moment generally larger in post-perovskite than in the liquid.

We tested our  free-spin treatment against fixed-spin-moment calculations. We minimized the free energy of iron bearing systems with respect to the imposed net cell magnetization \cite{edgington2019top}.  We find, at 5.63~g~cm$^{-3}$ and 6000~K, substitution free energies of $-2.31\pm0.04$~eV for post-perovskite and $-3.79\pm0.11$~eV for the liquid, in agreement with the corresponding free-spin values of $-2.33\pm0.08$ and $-3.75\pm0.15$~eV (Fig.~\ref{figS1}b). Thus, although the free-spin procedure underestimates the magnetic moment by 0.7-0.9 $\mu_B$ (Fig.~\ref{figS1}c), it reproduces the substitution free energies that control partitioning. Additional details of the spin-state comparison are shown in Fig.~\ref{figS1}.

\subsection{Equations of state}
\label{sec:methods-eos}

We obtain equations of state for iron-free MgSiO$_3$ post-perovskite and liquid from simulations along the 4000, 6000, 8000, and 10000~K isotherms. %
The raw simulation data of each phase are fitted to a thermal equation of state consisting of a third-order Birch--Murnaghan ``cold'' isotherm at 4000 K and a Mie--Gr\"uneisen thermal-pressure contribution,
\begin{equation}
P(V,T)=P_{\mathrm{0}}(V)+P_{\mathrm{th}}(V,T).
\label{eq:eos}
\end{equation}
The reference 4000 K isotherm is
\begin{equation}
P_{\mathrm{0}}(V) = \frac{3}{2}K_0
\left(x^{7/2}-x^{5/2}\right)
\left[1+\frac{3}{4}(K_0'-4)(x-1)\right],
\label{eq:eos-cold}
\end{equation}
where \(x=\left(V_0/V\right)^{2/3}\), and the thermal-pressure contribution is
\begin{equation}
P_{\mathrm{th}}(V,T)=\gamma_0
\left(\frac{V}{V_0}\right)^q
\frac{3Nk_B(T-T_0)}{V},
\label{eq:eos-thermal}
\end{equation}
where $N$ and $V$ are the number of atoms and volume of the simulation cell, respectively. The five parameters $(V_0,K_0,K_0',\gamma_0,q)$ are fitted simultaneously to all $(V,P,T)$ points for each phase, with the results shown in Table~\ref{tab:eos}. %

\begin{table}[t]
\centering
\begin{tabular}{l c c}
\hline
 & Post-perovskite & Liquid \\
\hline
$V_{0}$ (\AA$^{3}$/MgSiO$_3$) & $44.55 \pm 0.19$ & $57.09 \pm 0.95$ \\
$K_{0}$ (GPa) & $189.0 \pm 3.5$ & $75.8 \pm 6.2$ \\
$K_{0}'$ & $3.938 \pm 0.009$ & $4.083 \pm 0.053$ \\
$\gamma_{0}$ & $1.35 \pm 0.11$ & $1.28 \pm 0.17$ \\
$q$ & $0.36 \pm 0.12$ & $-0.35 \pm 0.17$ \\
\hline
\end{tabular}
\caption{Global thermal equation of state parameters (Eq.~\ref{eq:eos}) for MgSiO$_{3}$ post-perovskite and liquid, with reference temperature $T_{0} = 4000$~K. Uncertainties are 1$\sigma$ from the fit covariance ($n = 22$ points per phase). The fits are valid over the range 100-900 GPa.  
}
\label{tab:eos}
\end{table}

\begin{figure}[t]
\centering
\includegraphics[width=0.7\textwidth]{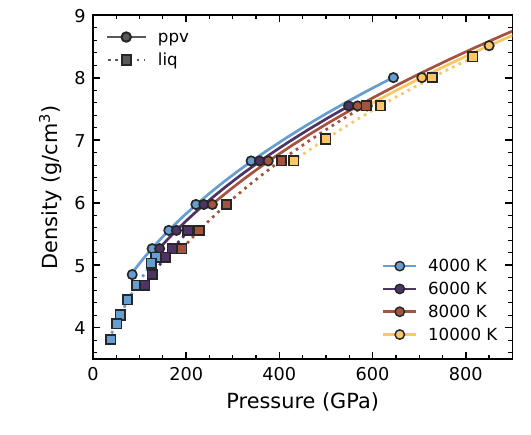}
\caption{Equation of state of MgSiO$_{3}$ post-perovskite (circles, solid curves) and MgSiO$_{3}$ liquid (squares, dotted curves) from the first-principles molecular dynamics simulations of this study. Symbols are the simulated state points, colored by temperature; curves are the global thermal equation of state (Eq.~\ref{eq:eos}, Table~\ref{tab:eos}) evaluated at the four isotherms.}
\label{fig1}
\end{figure}

Figure~\ref{fig1} shows the resulting isothermal equations of state of both MgSiO$_{\mathrm{3}}$ phases. At the same pressure and temperature, post-perovskite is denser than the liquid throughout the range explored, but the difference narrows with increasing compression. Along the theoretical MgSiO$_{3}$ melting curve of \cite{deng2023melting}, the solid--liquid density difference decreases from 6.9\% at 100~GPa to 1.6\% at 600~GPa. This shrinking density contrast makes the buoyancy of the crystal increasingly sensitive to compositional differences generated by Fe partitioning. The two phases also differ in the sign of $q$: the Gr\"uneisen parameter of post-perovskite decreases on compression while that of the liquid increases, consistent with the general behavior of silicate liquids \cite{stixrude2009thermodynamics}.

Our equation of state of post-perovskite agrees with the available experimental data. The most direct comparison is with shock compression  of bridgmanite starting material \cite{fei2021melting}. Across the four solid states in that study, between 296 and 500 GPa (with calculated Hugoniot temperatures of 4180--9430 K), our equation of state reproduces the measured densities to within 0.6--1.8\%. For example, at 422 GPa and 7720 K, we obtain 6.93 g/cm$^{3}$ against the measured $6.88 \pm 0.09$~g/cm$^{3}$. Static compression of post-perovskite in the laser-heated diamond-anvil cell \cite{sakai2016experimental,komabayashi2008simultaneous} provides a comparison at lower temperature: densities along the 300~K compression curve of \cite{sakai2016experimental} are  3--4\% greater than ours along the 4000~K isotherm at 150--250~GPa, consistent with the expected thermal expansion between the two temperatures.

The liquid equation of state is likewise consistent with the shock data. For the molten Hugoniot state at 715 GPa of \cite{fei2021melting}, with a computed temperature of 11800 K, our fit predicts 7.83~g~cm$^{-3}$ (with a small extrapolation in temperature) against the measured $7.76 \pm 0.08$~g~cm$^{-3}$. Our liquid results also agree, within uncertainty, with the liquid Hugoniot densities of shocked enstatite at 243-433 GPa \cite{fratanduono2018thermodynamic}, where Hugoniot temperatures were measured directly by optical pyrometry.

\subsection{Interior structure of super-Earths}
\label{sec:methods-interior}

The core--mantle boundary (CMB) pressures of the model planets considered below are estimated from interior structure calculations of their radial profiles. These planets have two layers (mantle and core), with an Earth-like core mass fraction of 0.32. Both mantle and core are assumed to have adiabatic temperature profiles. We solve for hydrostatic equilibrium coupled with the Adams--Williamson equation and the adiabatic temperature gradient,
\begin{equation}
\frac{dP}{dr} = -\rho g,
\qquad
\frac{d\rho}{dr} = -\frac{\rho^{2} g}{K_{S}},
\qquad
\frac{dT}{dr} = -\frac{\rho g \gamma T}{K_{S}},
\qquad
g = \frac{G m(r)}{r^{2}},
\label{eq:interior}
\end{equation}
where $K_{S}$ is the adiabatic bulk modulus, $\gamma$ is the Gr\"uneisen parameter, and $m(r)$ is the mass enclosed within radius $r$. The mantle structure is computed with the thermodynamic code HeFESTo assuming a pyrolite composition and potential temperature of 1600 K \cite{stixrude2011thermodynamics,stixrude2024thermodynamics}. For the iron core, we use a Vinet equation of state with parameters from \cite{wicks2018crystal}. The temperature at the planet's center is chosen to match the iron melting temperature of \cite{kraus2022measuring}, to give a core that is entirely liquid. %
These calculations yield CMB pressures of 283, 425, and 575~GPa at 2, 3, and 4~$M_{\oplus}$, respectively.

\section{Results}
\label{sec:results}

\begin{figure}[t]
\centering
\includegraphics[width=0.7\textwidth]{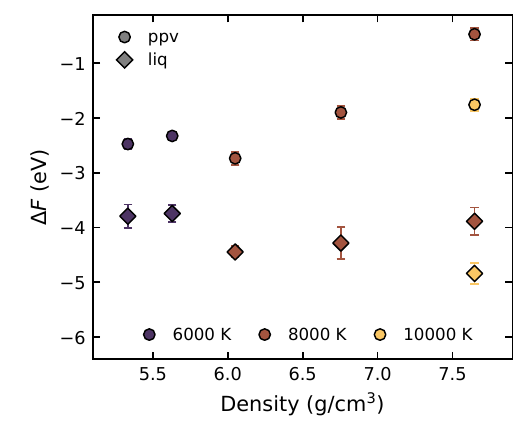}
\caption{Free energy of iron substitution $\Delta F$ vs density in post-perovskite (circles) and liquid (diamonds), as calculated by Eq. \ref{eq:TI}. Color denotes temperature. }
\label{fig2}
\end{figure}

\subsection{Distribution coefficient}
\label{sec:results-kd}

The free energy change of iron substitution computed by thermodynamic integration is shown in Fig.~\ref{fig2} for both phases. Iron prefers the liquid: $\Delta F$ is consistently lower (more negative) for the liquid than for post-perovskite at every condition explored. Upon compression, $\Delta F$ increases in both phases, but more steeply in post-perovskite: along the 8000~K isotherm, $\Delta F_{\mathrm{ppv}}$ increases by 2.3~eV between 6.0 and 7.7~g~cm$^{-3}$ while $\Delta F_{\mathrm{liq}}$ increases by only 0.5~eV. The free energy difference between the two phases---the quantity that sets $K_{D}$ through Eq.~\eqref{eq:KD}---therefore grows with compression: iron becomes increasingly incompatible in post-perovskite as pressure increases.

\begin{figure}[t]
\centering
\includegraphics[width=1\textwidth]{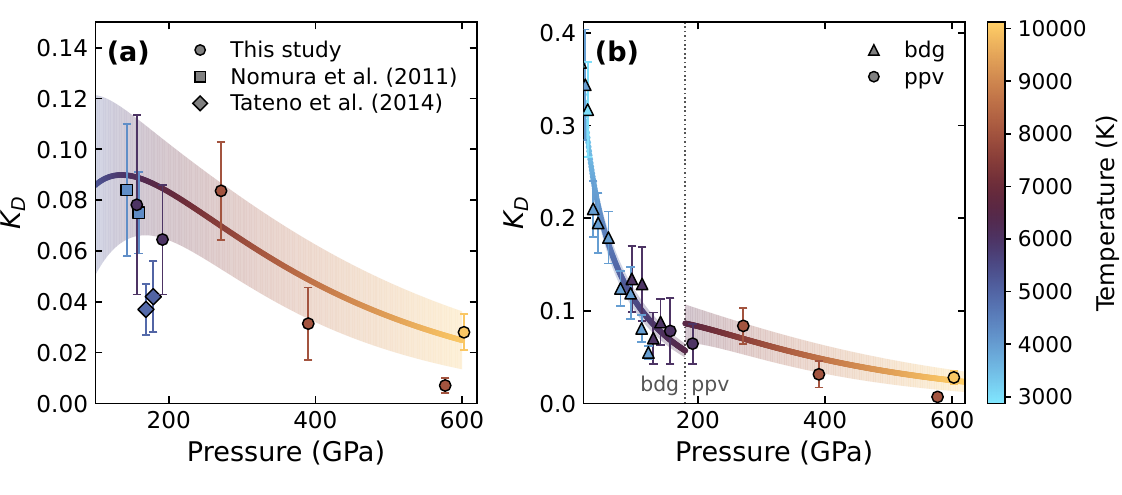}
\caption{Fe--Mg distribution coefficient $K_D$ vs pressure between the crystallizing silicate and coexisting melt. Symbols are colored by temperature. %
\textbf{(a)} Post-perovskite (circles, this study) compared with diamond-anvil cell experiments (squares, \cite{nomura2011spin}; diamonds, \cite{tateno2014melting}). The curve is the fit of Eq.~\eqref{eq:kdfit} evaluated along the post-perovskite melting curve of \cite{deng2023melting}.  %
\textbf{(b)} $K_{D}$ across the bridgmanite (triangles, \cite{dragulet2024partitioning}) to post-perovskite transition. Lines again represent fits evaluated along melting curves \cite{deng2023melting}, with the dotted vertical line representing the triple point at 180~GPa.%
}
\label{fig3}
\end{figure}

The distribution coefficient is significantly less than unity at all conditions explored and decreases gradually with increasing pressure (Fig.~\ref{fig3}). The temperature dependence is weaker; at fixed pressure, $K_{D}$ increases slightly with temperature. At the low pressure end, our results agree with the diamond-anvil cell measurements of \cite{nomura2011spin} and \cite{tateno2014melting}. We fit our simulated values to%
\begin{equation}
K_{D}(P,T) = \exp\!\left[-\frac{\Delta E - T\Delta S + P \Delta V}{k_{B}T}\right],
\label{eq:kdfit}
\end{equation}
so that $\Delta E$, $\Delta S$, and $\Delta V$ are respectively the energy, entropy, and volume of the exchange reaction \eqref{eq:rxn} per exchanged cation. A weighted least-squares fit to our six $K_{D}$ points (156--603~GPa, 6000--10000~K) yields $\Delta E = 122 \pm 64$~kJ/mol, $\Delta S = 12.9 \pm 11.2$~J/mol/K, and $\Delta V = 0.52 \pm 0.09$~cm$^{3}$/mol. A positive $\Delta V$ means that compression increasingly favors iron in the liquid relative to post-perovskite. The associated compression work becomes large at super-Earth pressures: $P\Delta V$ reaches $\sim$160~kJ/mol at 300~GPa, already exceeding the exchange energy $\Delta E$. Evaluated along the MgSiO$_{3}$ post-perovskite melting curve \cite{deng2023melting}, the fit shows $K_{D}$ decreasing from $0.090$ near 150~GPa to $0.025$ at 600~GPa (Fig.~\ref{fig3}a). %

Figure~\ref{fig3}b compares the post-perovskite results with our previous bridgmanite calculations \cite{dragulet2024partitioning}. At the bridgmanite--post-perovskite--melt triple point at 180~GPa \cite{deng2023melting}, the transition produces only a modest step in $K_D$, from $0.057 \pm 0.007$ in bridgmanite to $0.086 \pm 0.021$ in post-perovskite. %
The influence of pressure on partitioning is similar in the two structures; the exchange volume for post-perovskite, $0.52 \pm 0.09$~cm$^{3}$~mol$^{-1}$, is indistinguishable within uncertainty from the bridgmanite value of $0.53 \pm 0.05$~cm$^{3}$~mol$^{-1}$ \cite{dragulet2024partitioning}. %

\subsection{Crystal buoyancy and the stability of basal magma oceans}
\label{sec:results-buoyancy}

\begin{figure}[t]
\centering
\includegraphics[width=1\textwidth]{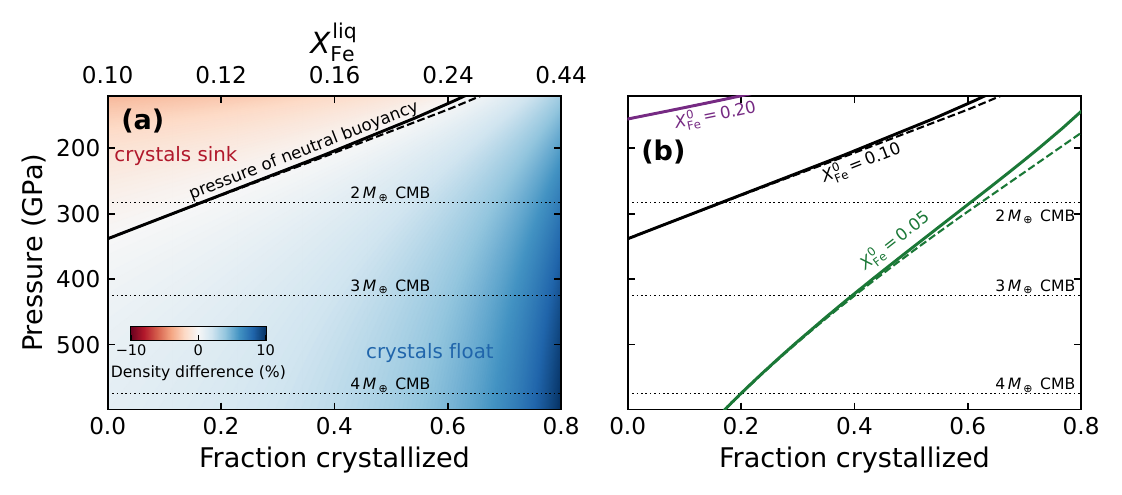}
\caption{Pressure vs fraction of (Mg,Fe)SiO$_{3}$ magma ocean that has crystallized. Dotted horizontal lines mark the CMB pressures of 2, 3, and 4~$M_{\oplus}$ planets. \textbf{(a)} density difference between post-perovskite and liquid (color) computed via fractional crystallization with starting bulk iron fraction of $X^{0}_{\mathrm{Fe}} = 0.10$. Bold black lines represent the pressure at which the densities of both phases are equal for fractional (solid) and batch (dashed) crystallization. 
\textbf{(b)} pressure of neutral buoyancy for varying bulk mantle iron fractions ($X^{0}_{\mathrm{Fe}} = 0.05,0.10,$ and $0.20$), denoted by color, for fractional (solid) and batch crystallization (dashed). Above each curve crystals sink; below it they float.
}
\label{fig4}
\end{figure}

We combine the fitted distribution coefficient with the equations of state of both phases along the MgSiO$_{\mathrm{3}}$ melting curve \cite{deng2023melting} to evaluate whether post-perovskite crystals float in super-Earth magma oceans (Fig.~\ref{fig4}). We consider a magma ocean of initial composition (Mg$_{1-X_{\mathrm{Fe}}^0}$,Fe$_{X_{\mathrm{Fe}}^0}$)SiO$_{3}$ that undergoes either fractional crystallization, described by
\begin{equation}
    \frac{dX_{\mathrm{Fe}}^{\mathrm{liq}}}{d \phi} = \left(D_{\mathrm{Fe}}-1 \right) \frac{X_{\mathrm{Fe}}^{\mathrm{liq}}}{\phi}
\end{equation}
or batch crystallization:
\begin{equation}
    X^{\mathrm{liq}}_{\mathrm{Fe}} = \frac{X^{\mathrm{0}}_{\mathrm{Fe}}}{\phi \left(1-D_{\mathrm{Fe}} \right) + D_{\mathrm{Fe}}}
\end{equation}
where \(D_{\mathrm{Fe}}=K_D\left(X^{\mathrm{ppv}}_{\mathrm{Mg}}/X^{\mathrm{liq}}_{\mathrm{Mg}}\right)\) is the iron partition coefficient and $\phi$ is the melt fraction. We then evaluate the density of the coexisting solid and liquid as functions of pressure and of the fraction of the ocean crystallized, $1-\phi$. We assume that volume varies linearly with iron content, and that $V_{\mathrm{FeSiO_{3}}}-V_{\mathrm{MgSiO_{3}}}=0.8$ cm$^3$/mol for the solid, and $0.6$ cm$^3$/mol for the liquid \cite{dragulet2024partitioning}. %

For an Earth-like bulk mantle iron fraction, $X_{\mathrm{Fe}}^0 = 0.10$, the pressure of neutral buoyancy, where solid and liquid densities are equal, lies at 340~GPa at the onset of crystallization and decreases as freezing proceeds (Fig.~\ref{fig4}a): iron enrichment of the residual liquid, shown on the top axis, is what drives the boundary upward. At pressures greater than the neutral-buoyancy pressure (i.e., deeper), crystals float; at lower pressures they sink.  Consequently, planets whose CMB pressures exceed the initial neutral-buoyancy pressure -- approximately those with mass exceeding 2.4~$M_{\oplus}$ -- host a gravitationally stable basal magma ocean from the moment crystallization begins: the first crystals to form at the base of the mantle are buoyant and rise, leaving a dense melt layer at the CMB. At lower CMB pressures ($< 340$ GPa), including that of a 2~$M_\oplus$ mantle, the first crystals are denser than the liquid, but progressive Fe enrichment lowers the neutral buoyancy pressure until newly formed post-perovskite becomes buoyant. Therefore, a BMO in such intermediate-mass planets may emerge only after a finite degree of solidification has driven the residual liquid to sufficiently iron-rich compositions. 

Bulk mantle compositions of rocky exoplanets are unknown, but they strongly affect the density crossover (Fig.~\ref{fig4}b). For a Mars-like mantle iron fraction of $X_{\mathrm{Fe}}^0=0.20$ \cite{bertka1997mineralogy}, the initial neutral-buoyancy pressure is only $\sim155$~GPa, whereas for $X_{\mathrm{Fe}}^0=0.05$ it lies at much greater pressure. Even in this latter iron-poor case, post-perovskite will become buoyant after some crystallization. Similar results are obtained with fractional and batch crystallization models, indicating that the buoyancy structure of a crystallizing super-Earth magma ocean is predominantly set by the partitioning and bulk composition, rather than by the crystallization mechanism.

\subsection{Liquidus depression of the residual liquid}
\label{sec:results-fpd}

\begin{figure}[t]
\centering
\includegraphics[width=0.7\textwidth]{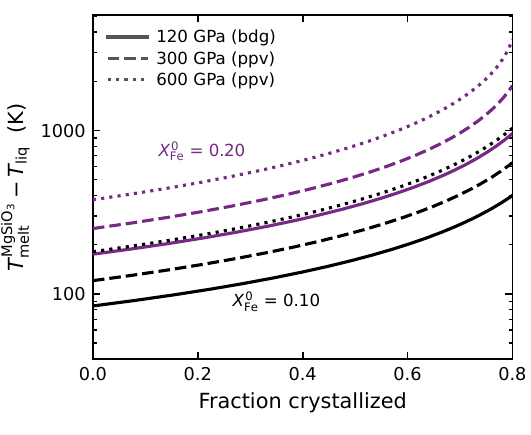}
\caption{Estimated liquidus depression of the residual melt during fractional crystallization at fixed pressure, for bulk iron contents $X_{\mathrm{Fe}}^0=0.10$ (black) and 0.20 (purple). At 120~GPa (solid curves) the crystallizing phase is bridgmanite; at 300~GPa (dashed) and 600~GPa (dotted) it is post-perovskite, each with its own $K_{D}$ (Fig.~\ref{fig3}). The depression is calculated from the MgSiO$_{3}$ melting temperature at the same pressure \cite{deng2023melting} -  see Fig.~\ref{figS2}.}
\label{fig5}
\end{figure}

The same partitioning that modifies crystal buoyancy also lowers the melting temperature of the residual liquid. We estimate this effect by treating solid and liquid as ideal MgSiO$_3$--FeSiO$_3$ solutions \cite{dragulet2024partitioning}. Equating the chemical potentials of the two phases and expanding the MgSiO$_3$ free energy of melting to first order about its melting temperature gives
\begin{equation}
T_{\mathrm{liq}}(P,X^{\mathrm{liq}}_{\mathrm{Fe}}) = \frac{T_{m}(P)} {1 + \dfrac{R}{\Delta S_{m}(P)} \left\{-\ln\left[1-\left(1-K_D\right)X^{\mathrm{liq}}_{\mathrm{Fe}}\right]\right\}},
\label{eq:liquidus}
\end{equation}
where $X^{\mathrm{liq}}_{\mathrm{Fe}}$ is the iron fraction of the liquid, $T_m(P)$ and $\Delta S_m(P)$ are the melting temperature and entropy of melting of MgSiO$_3$ \cite{deng2023melting}, and $K_D$ is evaluated at the liquidus temperature (Fig. \ref{fig3}). %
The liquidus depression as the liquid crystallizes is shown in Fig. \ref{fig5}, and the liquidus for various iron fractions is shown in Fig.~\ref{figS2}. The liquidus depression increases as crystallization enriches the residual liquid in iron and is larger at higher pressure. The initially gradual depression steepens markedly at advanced crystallization, reflecting both the nonlinear dependence of the liquidus on iron content and the rapid rise in iron content at later stages of freezing. For an initial iron fraction $X_{\mathrm{Fe}}^0=0.10$, the depression reaches several hundred kelvin over much of the crystallization interval and can approach $\sim1000$~K at advanced crystallization at 600~GPa. The effect is greater for $X_{\mathrm{Fe}}^0=0.20$. The pressure dependence predominantly reflects  the increasing incompatibility of iron on compression. %
This first-order estimate illustrates that the same iron enrichment responsible for stabilizing dense basal melt can significantly lower the liquidus, especially at elevated pressure, providing an additional thermodynamic effect that favors BMO persistence.

\section{Discussion}
\label{sec:discussion}

Our results extend the strong incompatibility of iron previously found in bridgmanite \cite{dragulet2024partitioning} to post-perovskite at the much higher pressures of super-Earth mantles. The nearly identical Fe--Mg exchange volumes in bridgmanite ($0.53 \pm 0.05$~cm$^{3}$~mol$^{-1}$) and post-perovskite ($0.52 \pm 0.09$~cm$^{3}$~mol$^{-1}$) show that the tendency toward greater incompatibility with increasing pressure survives the structural transition between the two phases. 

The primary planetary consequence is the effect of iron partitioning on crystal buoyancy. The density contrast between MgSiO$_3$ crystal and liquid is a few percent and decreases strongly with pressure, allowing preferential partitioning of iron into the liquid to reverse this contrast and make post-perovskite buoyant in the Fe-enriched melt (Fig.~\ref{fig4}). Our results show that post-perovskite flotation will occur throughout its stability field, either at the onset of crystallization (above 340 GPa for bulk mantle iron fraction of $X_{\mathrm{Fe}}^0=0.10$), or later on in the crystallization process after the liquid has become further enriched in iron. The depth of neutral buoyancy becomes shallower over time, potentially allowing a BMO to grow upward during solidification.  Iron enrichment further promotes the persistence of this reservoir by depressing its liquidus (Fig.~\ref{fig5}).

While the style of crystallization has a minimal effect on the location of neutral buoyancy, it will affect the dynamics and evolution of a crystallizing magma ocean \cite{caracas2019melt}. Recent calculations predict bridgmanite grain sizes on the order of 0.01 to 1 meters, sufficiently large for crystal--liquid segregation and fractional crystallization \cite{deng2026potential}. If post-perovskite behaves similarly, crystals should efficiently migrate according to their buoyancy, making the neutral-buoyancy pressure calculated here the primary control on the segregation of crystal and melt. 

Our crystal buoyancy calculations adopt the MgSiO$_3$ melting curve of \cite{deng2023melting}. A pair of dynamic compression studies that measure temperature along the Hugoniot find substantially lower melting temperatures at high pressure \cite{fratanduono2018thermodynamic, huff2026temperature}. Adopting the lower melting temperatures would  strengthen our conclusions because $\partial\ln K_D/\partial T>0$, implying smaller $K_D$ and therefore greater iron enrichment of the residual liquid. In addition, our calculations describe the MgSiO$_3$--FeSiO$_3$ binary only; additional components and phases may shift the location of neutral buoyancy. Nevertheless, interior structure and thermal models of rocky exoplanets should consider the effects of a basal magma ocean \cite{lherm2026magnetic}.

\section{Conclusions}
\label{sec:conclusions}

We used first-principles molecular dynamics and thermodynamic integration to determine Fe--Mg partitioning between post-perovskite and silicate liquid at 150--600~GPa and 6000--10,000~K, conditions spanning the deep mantles of rocky super-Earths. Iron is strongly incompatible in post-perovskite. At the low-pressure end of the post-perovskite stability field, we find $K_D$ in agreement with diamond-anvil cell experiments. Along the MgSiO$_3$ melting curve, $K_D$ decreases by nearly a factor of 4 from 150 to 600 GPa. The Fe--Mg exchange volume is indistinguishable from that previously found for bridgmanite, indicating that increasing iron incompatibility with pressure persists across the bridgmanite--post-perovskite transition.

Super-Earths are likely to contain long-lived basal magma oceans.  As post-perovskite crystallizes, iron accumulates in the residual liquid and increases its density, causing crystals to become buoyant either at the start of crystallization or after further crystallization enhances iron enrichment in the liquid. This same iron enrichment also depresses the liquidus, prolonging the crystallization timescale. These effects favor the segregation and retention of iron-rich melt at the base of rocky exoplanet mantles.  The presence of an iron-enriched basal magma ocean in super-Earths may be important for understanding the possible generation of magnetic fields in these bodies, as it is in the case of the early Earth \cite{dragulet2025electrical}.  %

\section*{Declaration of competing interest}
The authors have no competing interests to declare.

\section*{Acknowledgments}
This project was supported by the National Science Foundation under Grant EAR-2223935 to L.S. F.D. was supported by the Future Investigators in NASA Earth and Space Science and Technology under Grant 80NSSC24K1713. Calculations were carried out using the Hoffman2 Shared Cluster provided by UCLA Institute for Digital Research and Education's Research Technology Group.

\section*{Data availability}
Data will be made available on request.

\bibliographystyle{elsarticle-num}
\bibliography{references}

\clearpage
\section*{Supplementary material}
\setcounter{figure}{0}
\renewcommand{\thefigure}{S\arabic{figure}}

\begin{figure}[!h]
\centering
\includegraphics[width=\textwidth]{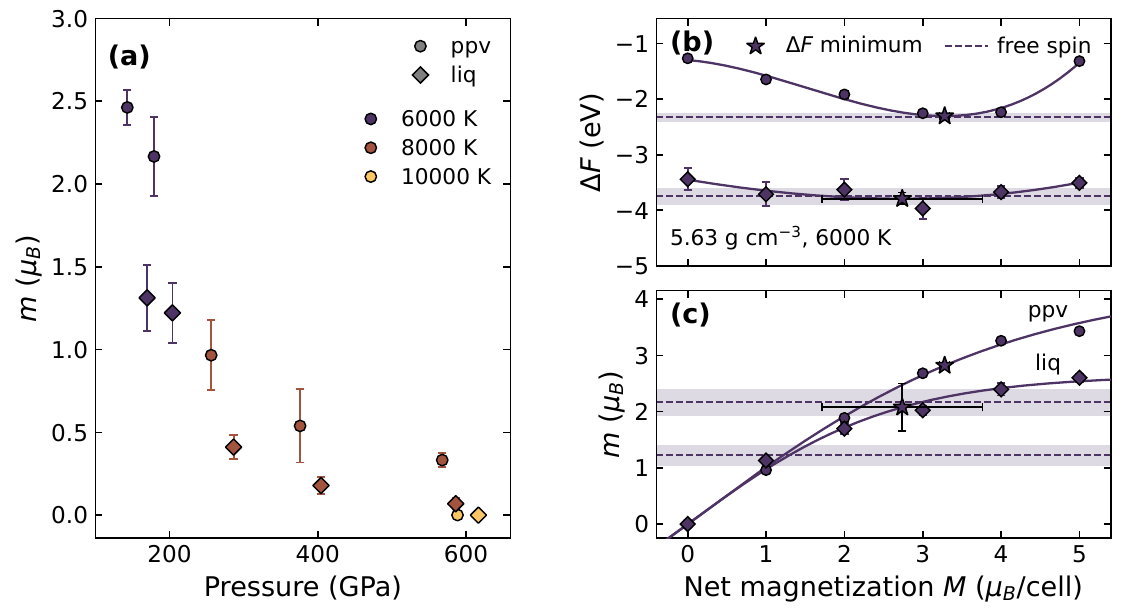}
\caption{Free-spin molecular dynamics recovers the free energy of iron substitution even though it understates the iron spin state. Circles denote post-perovskite and diamonds the liquid.
\textbf{(a)} Time-averaged local magnetic moment $m$ of Fe from unconstrained (free-spin) molecular dynamics simulations vs pressure, colored by temperature. 
\textbf{(b)} Free energy of iron substitution $\Delta F$ against the net cell magnetization $M$ imposed with the fixed-spin-moment constraint (NUPDOWN flag in VASP), at 5.63~g~cm$^{-3}$ and 6000~K ($\sim$190 GPa). Solid curves are weighted cubic fits and stars mark their minima; dashed lines and shaded bands are the free-spin $\Delta F$ and uncertainty. The constrained spin minima match the free-spin values well within uncertainty. 
\textbf{(c)} Local magnetic moment $m$ of Fe atom over the same constrained scan, with fits $m(M) = m_{\mathrm{sat}} \tanh(M/M_{0})$ (lines); stars give the moment at the $\Delta F$ minimum of panel (b), which exceeds the free-spin time average (dashed lines and bands) by 0.7--0.9~$\mu_{B}$.}
\label{figS1}
\end{figure}

\clearpage

\begin{figure}[!h]
\centering
\includegraphics[width=0.7\textwidth]{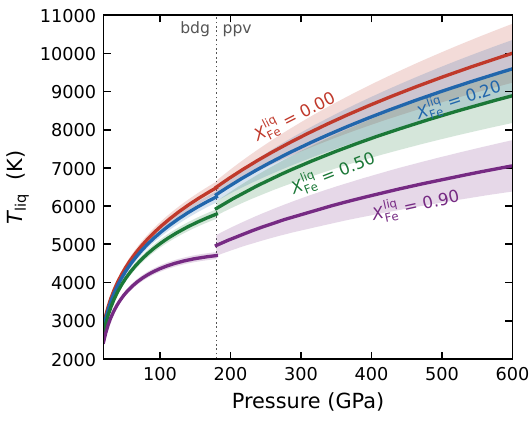}
\caption{Estimated liquidus temperature of the (Mg,Fe)SiO$_{3}$ liquid as a function of pressure for liquid iron fractions $X_{\mathrm{Fe}}^{\mathrm{liq}} = 0.00$ (MgSiO$_{3}$ melting curve \cite{deng2023melting}), 0.20, 0.50, and 0.90, with the crystallizing phase switching from bridgmanite to post-perovskite at the 180~GPa triple point \cite{deng2023melting} (dotted vertical line).%
}
\label{figS2}
\end{figure}

\end{document}